\documentclass{phbauth}
\usepackage{graphicx}

\begin{document}

\newcommand{\bv}[1]{\mbox{\boldmath$#1$}}
\begin{frontmatter}

% Use lower case letters in the title.
\title{
%% Vortices in Bose-Einstein condensate with spin degree of freedom
A simple method to create a vortex
in Bose-Einstein condensate of alkali atoms
}

\author[address1]{Mikio Nakahara\thanksref{thank1}},
\author[address2]{Tomoya Isoshima},
\author[address2]{Kazushige Machida},
\author[address1]{Shin-ichiro Ogawa},
\author[address3]{Tetsuo Ohmi}

\address[address1]{Department of Physics, Kinki University, Higashi-Osaka
577-8502, Japan}
\address[address2]{Department of Physics, Okayama University, Okayama
700-8530, Japan}
\address[address3]{Department of Physics, Kyoto University, Kyoto 601-8501,
Japan }

% The corresponding author should be distinguished and his email
% address and/or fax number must be given. His mailing address has to
% be complete: the proofs are send to this address around
% January 1, 2000. The address for sending proofs has to be indicated
% as "present address", if it is different from the address above.
\thanks[thank1]{Corresponding author. E-mail: nakahara@math.kindai.ac.jp,
Fax Number 81-6-6727-4301}
%\thanks[thank2]{Present address: Neverland, USA}

\begin{abstract}
Bose-Einstein condensation in alkali atoms has materialized quite
an interesting system, namely a condensate with a spin degree of freedom.
In analogy with the A-phase of the superfluid $^3$He, numerous textures
with nonvanishing vorticity have been proposed. In the present paper,
interesting properties of such spin textures are analyzed. 
We propose a remarkably simple method to create a vortex state of a BEC
in alkali atoms.
\end{abstract}

\end{frontmatter}

%\section{Introduction}
The discovery of Bose-Einstein condensation (BEC) in alkali atoms has
led to several exciting fields to study. One of these developments is a
BEC with a spin degree of freedom \cite{b1}. When
the spin exchange interaction is ferromagnetic or the confining
magnetic field is strong, the BEC behaves somewhat similarly to superfluid 
$^3$He-A, where topological objects called textures
are known to exist. In the present paper, we investigate vortices
in a BEC, which is analogous to the vortex
and the disgyration in $^3$He-A. We first explain the order parameter
of a BEC with a spin degree of freedom. Then the cross
disgyration, expected to exist in the Ioffe-Pritchard (IP) trap, is considered.
Finally, we propose a simple method to create an ordinary vortex line
in a BEC.
% by making use of this degree of freedom.

Suppose an alkali atom has a hyperfine-spin $|F|=1$.
Then the order parameter has three components $\Psi_{\pm 1}$ and $\Psi_0$,
which represent the amplitudes with $F_z=\pm 1, 0$, respectively.
The basis vectors in this representation are
$\{|\pm \rangle, |0 \rangle\}$. 
We introduce another set of basis vectors
$|x \rangle, |y \rangle$ and $|z \rangle$, which are defined by
$F_x |x \rangle= F_y |y \rangle=F_z |z \rangle=0$. These vectors
are related with the previous vectors as
$|\pm 1\rangle = \mp (1/\sqrt{2})\left( |x \rangle \pm i|y \rangle\right)$
and $|0 \rangle=|z \rangle$.
When the $z$-axis is taken parallel to the uniform magnetic field,
the order parameter of the weak field seeking state
takes the form $\Psi_{-1} = \psi$ and $\Psi_0=\Psi_{1}
=0$, which is also written as $\Psi_x=i\Psi_y=\psi/\sqrt{2}$. 
Let us denote this state in a vectorial form as
$\bv{\Psi} =(\psi/\sqrt{2}) \left(\hat{\bv{x}} -i \hat{\bv{y}}\right)$,
where the common factor has been absorbed in the amplitude $\psi$.
Suppose the magnetic field points a direction
$\hat{\bv{B}} =(\sin \beta \cos \alpha, \sin \beta \sin \alpha, \cos \beta)$.
Then the weak field seeking state takes the form
$\bv{\Psi} = (\psi/\sqrt{2}) e^{i \gamma} (\hat{\bv{m}} - i \hat{\bv{n}})$,
where
$\hat{\bv{m}} = (\cos \beta \cos \alpha, \cos \beta \sin \alpha, -\sin \beta)$
and $\hat{\bv{n}}= (- \sin \alpha, \cos \alpha, 0)$.
The unit vector $\hat{\bv{l}} =- \hat{\bv{m}} \times \hat{\bv{n}}$
is the direction of the spin polarization.
The same amplitudes in the basis $\{|0 \rangle, |\pm \rangle\}$ are \cite{b2}
\begin{eqnarray}
\Psi_1 &=& (\psi/2) (1 - \cos \beta) e^{-i \alpha + i \gamma} \nonumber\\
\Psi_0 &=& - (\psi/\sqrt{2}) \sin \beta e^{i \gamma}\\
\Psi_{-1} &=& (\psi/2) (1 + \cos \beta) e^{i \alpha + i \gamma} \nonumber
\end{eqnarray}

The choice $\alpha = -\phi, \beta = \pi/2$ yields the cross disgyration
shown in Fig.1, where $\phi$ is the azimuthal angle.
This texture has a nonvanishing 
vorticity
$n$ when
$\gamma = n \phi$. It is expected that the cross disgyration is
realized in the IP trap with $B_z=0$.
\begin{figure}[btp]
%h=here, t=top, b=bottom, p=separate figure page
\begin{center}\leavevmode
\includegraphics[width=0.5\linewidth]{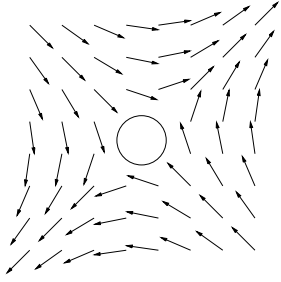}
\caption{
The $\hat{l}$-vector field of a cross disgyration. The circle shows
the laser beam, where no atoms exist.
}\label{figurename}\end{center}\end{figure}

Suppose a strong magnetic field $B_z$
is applied to a BEC, along with the quadrapole field.
Atoms are assumed to be in the weak field seeking state
so that they are confined in the trap. The field
$B_z$ is so strong compared to the quadrapole field that the order parameter
is virtually
$\bv{\Psi} = (\psi/\sqrt{2})(\hat{\bv{x}} - i \hat{\bv{y}})$.
 Clearly the vorticity of this texture vanishes.
This configuration is derived from Eq.(1) by putting $\beta=0$
and $\gamma=-\alpha=\phi$. Then $B_z$ is
adiabatically decreased, so that $\hat{\bv{l}}$ is always antiparallel to
$\bv{B}$, until $B_z$ vanishes.
The adiabatic condition is required for atoms to
remain in the weak field seeking state.
Then the cross disgyration appears in the presence of the quadrapole field.
In due process, the angle $\beta$ increases from $0$
to $\pi/2$ and $\gamma$ and $-\alpha$ are identified with $\phi$.
Here the trap must be plugged by laser beams along the axis, the top
and the bottom of the trap so that the atoms do not escape from the trap.
In the final step, the external field $B_z$ is gradually increased
in the opposite ($-z$) direction. Then the $\hat{\bv{l}}$-vector
points up so that $\beta= \pi$. Substituting these angles into (1), one obtains
\begin{equation}
\Psi_{-1} = \Psi_0 = 0, \quad
\Psi_{1}  =  \psi e^{2i \phi}.
\end{equation}
%
%----
% This is nothing but the order parameter of a vortex with vorticity 2.
%-
This is nothing but the order parameter of a vortex with the winding number 2.
%----
The amplitude $\psi(r)$ is determined by solving the Gross-Pitaevskii
equation with appropriate boundary conditions.
The supercurrent of the texture has the $\phi$-component
\begin{equation}
j_{s\phi} = m |\psi|^2 v_{s\phi},\ 
v_{s\phi} = (\hbar/mr) (1-\cos \beta),
\end{equation}
where $m$ is the atomic mass and $v_s$ is the superfluid velocity.
If the Na mass is substituted into Eq.(3), we obtain $v_{s\phi}
\simeq 0.5/r\ {\rm cm/s}$ 
for $\beta=\pi$, where $r$ is measured in units of $\mu {\rm m}$.

In summary, we have proposed a simple method to create
a vortex in a BEC of alkali atoms. A strong magnetic field
is applied along the axis of the IP trap field, which is
adiabatically varied toward a negative large value.
%--------
% Starting with a uniform order parameter field, we are eventually
% left with a vortex of the strength 2.
%-
Starting with a uniform order parameter field, we are eventually
left with a vortex of the winding number 2.
%--------
The initial state has no
circulation while the final state does. This is because the external
magnetic field transfers torque to the BEC while the spin vector is
turned upside down. 
%-----
%% When the hyperfine-spin is $F$ in general,  we will end up with a vortex
%% with the winding number $2F$ since $\Psi_{-F}$ has a phase
%% $e^{iF(\alpha+\gamma)}$. 
When the hyperfine-spin is $F$ in general,
we will end up with a vortex
with the winding number $2F$
since
$\Psi_{-F}$ and $\Psi_{F}$ have phases 
$F(\alpha+\gamma)$ and $F(-\alpha+\gamma)$ respectively. 
%-----

% %----------------
% \begin{ack}
We would like to thank M.~Mitsunaga, Y.~Takahashi and T.~Yabusaki
for discussions. The work of MN is supported partially by the
Grant-in-Aid for Scientific Research Fund of the Ministry of Education,
Science, Sports and Culture, No.11640361.
% \end{ack}
% %----------------

\end{document}